\documentclass[pra,aps,showpacs,superscriptaddress,twocolumn,10pt]{revtex4-1}
\usepackage{amsfonts}
\usepackage{amsmath}
\usepackage{amssymb}
\usepackage{graphicx}
\usepackage{graphics}
\usepackage{epsfig}
\usepackage[utf8]{inputenc}
\usepackage{lmodern}
\usepackage[normalem]{ulem}
\usepackage{color}
\usepackage[colorlinks]{hyperref}

\hypersetup{colorlinks, citecolor=blue, linkcolor=red, urlcolor=blue}

\newcommand{\orcid}[1]{\href{https://orcid.org/#1}{\includegraphics[width=7pt]{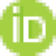}}}

\begin{document}

\title{NMR Hamiltonian as an effective Hamiltonian to generate Schr\"{o}dinger's cat states}

\author{A. Consuelo-Leal\orcid{0000-0003-1141-210X}}\email{adrianeleal@ifsc.usp.br}
\affiliation{Instituto de F\'{i}sica de S\~{a}o Carlos, Universidade de S\~{a}o Paulo, CP 369, 13560-970, S\~{a}o Carlos, S\~{a}o Paulo, Brasil.}

\author{A. G. Araujo-Ferreira\orcid{0000-0002-6676-384X}}
\affiliation{Instituto de F\'{i}sica de S\~{a}o Carlos, Universidade de S\~{a}o Paulo, CP 369, 13560-970, S\~{a}o Carlos, S\~{a}o Paulo, Brasil.}

\author{E. L. G. Vidoto\orcid{0000-0002-6876-3309}}
\affiliation{Instituto de F\'{i}sica de S\~{a}o Carlos, Universidade de S\~{a}o Paulo, CP 369, 13560-970, S\~{a}o Carlos, S\~{a}o Paulo, Brasil.}

\author{E. Lucas-Oliveira\orcid{0000-0003-1353-918X}}
\affiliation{Instituto de F\'{i}sica de S\~{a}o Carlos, Universidade de S\~{a}o Paulo, CP 369, 13560-970, S\~{a}o Carlos, S\~{a}o Paulo, Brasil.}

\author{T. J. Bonagamba\orcid{0000-0001-8894-9170}}
\affiliation{Instituto de F\'{i}sica de S\~{a}o Carlos, Universidade de S\~{a}o Paulo, CP 369, 13560-970, S\~{a}o Carlos, S\~{a}o Paulo, Brasil.}

\author{R. Auccaise\orcid{0000-0002-9602-6533}}\email{raestrada@uepg.br}
\affiliation{Departamento de F\'{i}sica, Universidade Estadual de Ponta Grossa, Av. Carlos Cavalcanti 4748, 84030-900 Ponta Grossa, Paran\'{a}, Brasil.}

\date{\today}

\begin{abstract}
This report experimentally demonstrates that the theoretical background of the atom-field scenario  points out that the NMR quadrupolar Hamiltonian works as an effective Hamiltonian to generate  Schr\"{o}dinger's cat states in a $2I+1$ low dimensional Hilbert space. The versatility of this nuclear spin setup is verified by monitoring the $^{23}$Na nucleus of a lyotropic liquid crystal sample at the nematic phase. The  quantum state tomography and the Wigner quasiprobability distribution function are performed to characterize the accuracy  of the experimental implementation.
\end{abstract}


\maketitle

\section{Introduction}

The Schr\"{o}dinger's cat state is one of the most intriguing theoretical concepts that allows  establishing the frontiers between the classical and the quantum world \cite{buzek1995Book,agarwal2013Book}. This kind of quantum state has played an essential role in quantum computation. Theoretically, it is used as a resource for elucidating  the foundations of quantum mechanics  and, experimentally, it is a challenging task to implement.  Accordingly, promising  quantum experimental techniques started as a workbench with few particles such as  one trapped ion \cite{monroe1996} and one  pair of photons \citep{bouwmeester1997} were used to  highlight their quantum capabilities on the implementation of Bell states. Hence, from the point of view of quantum control, increasing the size and number of qubits to implement the Schr\"{o}dinger's cat state was a critical issue  for experimentalists. It took almost ten years for the first implementation with six \citep{leibfried2005}  trapped ions in a cavity at ultracold temperatures. Eight years later, reports came from implementations with dozens of photons \citep{deleglise2008,vlastakis2013} superconducting devices with five qubits \citep{leek2013}. The advantages of these quantum states open new possibilities for applications on quantum information procedures such as quantum simulation \citep{barreiro2011} and development on quantum technologies \citep{tiecke2014}.

The theoretical background on the implementation of Schr\"{o}dinger's cat state explores the atom-field interaction primarily \citep{deleglise2008,vlastakis2013}, but it is not the only possible strategy. In atomic physics, there is a theoretical proposal, which explores the atom-atom interaction \citep{agarwal1997,zheng2001,klimov1998,klimov2002JOB,james2000,prado2011}. This strategy was developed using the angular momentum description $\hat{\mathbf{J}}=\left( \hat{\mathbf{J}}_{x}, \hat{\mathbf{J}}_{y}, \hat{\mathbf{J}}_{z}\right)$, from the SU(2) algebra and the appropriate coupling strength between pairs of atoms of a two-mode Bose-Einstein Condensate. From  an algebraic structure's point of view, the nuclear spin angular momentum description, $\hat{\mathbf{I}}=\left( \hat{\mathbf{I}}_{x}, \hat{\mathbf{I}}_{y}, \hat{\mathbf{I}}_{z} \right)$,  belongs and obeys the SU(2) algebra, and this particular characteristic allows transferring the knowledge developed for a many-body system into a spin system \cite{auccaise2015}. In this work, we will show how a nuclear spin system, $I=3/2$, achieves an analogous behavior of a few ultra-cold atoms particles, $N=3$, in a trap.

In this sense, the introduction of atom-field definitions and their description in terms of nuclear spin notation is organized as follows. First, in Sec. \ref{sec:Theory}, the main theoretical background of the atom-field description to generate the  Schr\"{o}dinger's cat states is discussed and the corresponding definitions of a nuclear spin system,  and the Wigner quasiprobability distribution functions are presented. Next, in Sec. \ref{sec:DescriptionExperimentalProcedures}, the soft matter experimental setup for the nuclear spin system, the initialization of the quantum state, and the quantum state tomography procedure are briefly commented.  Then, in Sec.  \ref{sec:ExperimentalResults},  the experimental results generated by implementation of the  Schr\"{o}dinger's cat state are  detailed. Finally, the main results are discussed in Sec. \ref{sec:Discussions}, and the conclusions  are summarized in Sec. \ref{sec:Conclusions}.

\section{Theory}
\label{sec:Theory} 

\subsection{Atom description}
\label{sec:AtomDescription} 

This theoretical background section starts describing the quantum system used in this study: consider an ensemble of N identical two-level atoms interacting collectively with a single mode electromagnetic field \cite{agarwal1997,klimov2002JOB}, such that the electromagnetic field, when appropriately described, induces an atom system dynamics  to prepare them into a quantum superposition of coherent states of the type  $ \left\vert \zeta \right\rangle +\left\vert -\zeta \right\rangle  $  (avoiding the normalization constant), which is different from the GHZ type  \cite{gao2010, song2017PRL, cappellaro2005}. We applied three theoretical quantum mechanical methods to achieve the effective Hamiltonian  and we detailed their main assumptions in the following paragraphs.

{\textit{System-Reservoir quantum system:}} The first method considers an open quantum system approach in which a quantum field is described as a thermal bath inducing accurate dynamics of the atoms system to achieve the target quantum state \cite{agarwal1997,zheng2001}. In this sense the master equation  as denoted by Eq. (1-4) of Ref. \cite{agarwal1997} is represented explicitly by 
\begin{eqnarray*}
\frac{d\hat{\boldsymbol{\rho }}}{dt} &=&-ig\left[ \left( \hat{\mathbf{J}}_{+} \hat{\boldsymbol{a}} + \hat{\mathbf{J}}_{-} \hat{\boldsymbol{a}}^{\dag } \right) ,\hat{\boldsymbol{\rho }} \right] -i  \left( \omega _{c}- \omega \right)   \left[ \hat{\boldsymbol{a}}^{\dag }\hat{\boldsymbol{a}},\hat{\boldsymbol{\rho }} \right]  \\
&&+\kappa \left( \overline{n}+1\right) \left( 2\hat{\boldsymbol{a}}\hat{\boldsymbol{\rho }} \hat{\boldsymbol{a}}^{\dag } - \hat{\boldsymbol{a}}^{\dag } \hat{\boldsymbol{a}} \hat{\boldsymbol{\rho }} - \hat{\boldsymbol{\rho }} \hat{\boldsymbol{a}}^{\dag }\hat{\boldsymbol{a}}\right)  \\
&&+\kappa \left( \overline{n}\right) \left( 2\hat{\boldsymbol{a}}^{\dag }\hat{\boldsymbol{\rho }} \hat{\boldsymbol{a}} - \hat{\boldsymbol{a}} \hat{\boldsymbol{a}}^{\dag } \hat{\boldsymbol{\rho }} + \hat{\boldsymbol{\rho } }\hat{\boldsymbol{a}} \hat{\boldsymbol{a}}^{\dag } \right) \text{,}
\end{eqnarray*}
where  $g$ is the atom-field coupling constant, $\omega _{c} $ is the atom characteristic frequency, $\omega _{0} $ is the electromagnetic field characteristic frequency, $\kappa$ is the rate of the loss of photons and $\overline{n}$ is the average number of thermal photons in the cavity. The density operator $\hat{\boldsymbol{\rho }} \equiv \hat{\boldsymbol{\rho }}_{a} \otimes  \hat{\boldsymbol{\rho }}_{f} $ is the tensor product of the density operator of the atom $\hat{\boldsymbol{\rho }}_{a}$ and field $\hat{\boldsymbol{\rho }}_{f}$ system. Considering the field quantum state of the thermal equilibrium as denoted by Eq. (6) of the Ref. \cite{agarwal1997};  computing the definition of partial trace on the field degrees of freedom; and assuming the dynamics of the atom system evolves slowly when compared to the field system then the master equation will be rewritten as in Eq. (10) of Ref. \cite{agarwal1997}
\begin{equation*}
\frac{d\hat{\boldsymbol{\rho }}_{a}}{dt}=-i\frac{g^{2}\delta _{c}}{\kappa^{2}+\delta
_{c}^{2}}\left[ \hat{\mathbf{J}}_{+}\hat{\mathbf{J}}_{-}+2\overline{n}\hat{%
\mathbf{J}}_{z},\hat{\boldsymbol{\rho }}_{a}\right] 
\end{equation*}
where the effective Hamiltonian is the first component of the commutator operator, such that  using the properties of angular momentum operators (see Eq. (C-15-b) on Pag. 649 of Ref. \cite{cohen1977Book}) the effective Hamiltonian obeys the quadratic dependence on the $\hat{\mathbf{J}}_{z}$ as denoted in Eq. (\ref{HamiltonianoEffetivo}) of this manuscript.

{\textit{Unitary transformation and the Dicke model:}} The second method discusses the application of a unitary transformation approach such that the overall operators are denoted in an equivalent representation that favors the generation of the target quantum state  \cite{klimov1998,klimov2002JOB}. For example, see the general unitary transformation of Eq.(2.5) of Ref.  \cite{klimov2002JOB} valid for two coupled quantum systems, field-field or atom-field quantum system. In particular, we are interested on the atom-field quantum system as discussed in sec. 5 of Ref.  \cite{klimov2002JOB} where the Hamiltonian of Eq. (5.2) using the operators denoted in Eq. (5.3) of Ref.  \cite{klimov2002JOB} such that the  Hamiltonian of Eq. (5.6) is generated
\begin{equation*}
\hat{\mathcal{H}}_{\text{eff}}^{u.t.} \simeq \hbar \Delta \hat{\mathbf{J}}_{z}+\hbar \frac{g^{2}}{ \Delta }\left( \left( 2\hat{\boldsymbol{a}}^{\dag }\hat{\boldsymbol{a}}+\hat{\mathbf{1}}\right) \hat{\mathbf{J}}_{z}  - \hat{\mathbf{J}}_{z}^{2} + \hat{\mathbf{J}}^{2} \right) 
\text{,}
\end{equation*}
where $ \Delta $ is the detuning parameter between the characteristic frequency of the atoms and the field,  and $ \hbar $ is the reduced Planck's constant. Considering an additional rotating frame with $\omega_{r.f.}=\Delta$ and computing the partial trace on the field degree of freedom considering the field quantum state of the thermal equilibrium as denoted by Eq. (6) of the Ref. \cite{agarwal1997};  then the effective Hamiltonian of Eq. (\ref{HamiltonianoEffetivo}) is achieved.

{\textit{Highly detuned interactions:}} The third method applies the highly oscillating energetic contribution approach to generate the quantum state  \cite{james2000,prado2011}. In this case the method introduced by James et. al.  at the Eq. (21) of Ref.  \cite{james2000} is applied on a two-mode field system interacting with an ensemble of N identical neutral atom system as denoted by the Hamiltonian of Eq. (19) in Ref.  \cite{prado2011} and explicitly represented
\begin{widetext}
\begin{equation*}
\hat{\mathcal{H}}_{\text{eff}}^{h.d.i.} = +\epsilon \hat{\mathbf{J}}_{+}\hat{\mathbf{J}}_{-}+\left( \frac{\left\vert \lambda \right\vert ^{2}}{\Theta \hbar }2\hat{\mathbf{J}} _{+} \hat{\mathbf{J}}_{-} + \frac{\left\vert \lambda \right\vert ^{2}}{\Theta \hbar }\left( \hat{\boldsymbol{a}}^{\dag }\hat{\boldsymbol{a}}+\hat{\boldsymbol{a}}^{\dag }\hat{\boldsymbol{b}}+\hat{\boldsymbol{b}}^{\dag }%
\hat{\boldsymbol{a}}+\hat{\boldsymbol{b}}^{\dag }\hat{\boldsymbol{b}}\right) 2\hat{\mathbf{J}}_{z}\right) +\Omega 
\hat{\mathbf{J}}_{+}+\Omega ^{\ast }\hat{\mathbf{J}}_{-}\text{.}
\end{equation*}
\end{widetext}
and using the properties of raising and lowering operators from the fundamentals
of Quantum Mechanics (see Eq. (C-15-b) on page 649 of Ref. \cite%
{cohen1977Book}), applying the partial trace definition on the field degrees of freedom
 at the thermal equilibrium state where the mean number of photons at
each mode satisfies $\overline{n}_{a}=\overline{n}_{b}=\overline{n}$, neglecting the classical driving field on the two-level atoms ($\Omega=0$),   and executing the method as denoted by the Eq. (21) of Ref. \cite{james2000} the effective Hamiltonian can be generated by
\begin{widetext}
\begin{equation*}
\hat{\mathcal{H}}_{\text{eff}}^{h.d.i.} =  + \frac{\left\vert \lambda \right\vert ^{2}}{\Theta \hbar }\left( \frac{\Theta \hbar }{\left\vert \lambda \right\vert ^{2}}\epsilon +2+4 \overline{n}\right) \hat{\mathbf{J}}_{z}
- \left( \epsilon +\frac{ 2\left\vert \lambda \right\vert ^{2}}{\Theta \hbar }\right) \hat{\mathbf{J}} _ {z}^{2}
 +\left( \epsilon +\frac{2\left\vert \lambda \right\vert ^{2} }{\Theta \hbar }\right) \hat{\mathbf{J}}^{2} \text{,}
\end{equation*}
\end{widetext}
so that assigning the parameter $\epsilon =0$ and isolating the common factor $\frac{2\left\vert \lambda \right\vert ^{2}}{\Theta \hbar }$ of these three terms then the effective Hamiltonian of Eq. (\ref{HamiltonianoEffetivo}) is achieved.

Therefore, the two common mathematical assumptions must be itemized: (i) the rotating wave approximation, and (ii) the partial trace on the field degrees of freedom   to mathematically rewrite any atom-field Hamiltonian in terms of angular momentum operators that represent the atoms or particles of the system. Under these assumptions and the discussion of the three approaches, an effective Hamiltonian is established by \cite{agarwal1997,klimov1998}
\begin{equation}
\hat{\mathcal{H}}_{\text{eff}} \equiv \hbar \kappa \left(    \left( 2 \ \overline{n} + 1 \right)     \hat{\mathbf{J}} _{z} - \hat{\mathbf{J}}_{z}^{2} +  \hat{\mathbf{J}}^{2}\right)  \text{,}   \label{HamiltonianoEffetivo}
\end{equation}
 where   $\kappa$  is the   effective coupling constant.

Similarly to the theoretical and experimental arguments developed for the case of the atom-field approach, and from a functional mathematical point of view, the Hamiltonian of Eq. (\ref{HamiltonianoEffetivo}) fits the definition of a nuclear quadrupolar Hamiltonian \cite{slichter1992Book,wasylishen2012Book}. From their similar qualities, the nuclear spin counterpart has been explored in many quantum information applications, such as the quantum simulation of a Bose-Einstein condensate of one mode \cite{auccaise2009}, the definition of coherent states in the context of nuclear spin \cite{auccaise2013}, and others  \cite{araujo-ferreira2013,nie2015,auccaise2015,teles2018}.

\subsection{Nuclear spin description}
\label{sec:NuclearSpinDescription}

Any quadrupolar nucleus has a spin value $I>1/2$. Thus, any system composed of these nuclei is associated with a Hilbert space of dimension $d=2I+1$ generated by the Dicke basis of quantum states, denoted by  $\left\{ \left\vert I,m\right\rangle \right\} $ with  $ m=I,I-1,\ldots ,-I+1,-I$. The set  $\left\{ \left\vert I,m\right\rangle \right\} $  is an appropriate basis of eigenstates of the angular momentum operators $\hat{\mathbf{I}}_{z}$ and $\hat{\mathbf{I}}^{2}$ for the system, which satisfies $\hat{\mathbf{I}} _{z} \left\vert I,m\right\rangle= m  \left\vert I,m\right\rangle $ and $\hat{\mathbf{I}}^{2}   \left\vert I,m\right\rangle= I\left(  I+1\right)   \left\vert I,m\right\rangle $. In order to quantify the energy of the system, the Hamiltonian at the laboratory frame is defined with three main energy contributions. The first one,  the highest energetic contribution, is the Zeeman term which is the interaction between the magnetic moment of the nucleus and a strong static magnetic field along the $z$-axis, $ \hbar  \gamma B_{0} \hat{\mathbf{I}} _{z} $, where  $\gamma$ is the gyromagnetic ratio of the nucleus and  $B_{0}$ is the intensity of the static magnetic field. The second one, the quadrupolar term, is the interaction between the quadrupole moment of the nucleus and an electric field gradient around the nucleus with an anisotropy along the $z$-axis, $ \hbar \frac{\omega _{Q}}{6}\left( 3\hat{\mathbf{I}}_{z}^{2}-\hat{\mathbf{I}}^{2}  \right)  $, where  $\omega _{Q}$ is the quadrupolar angular frequency of the system  (the non-linear angular momentum operator dependence of this term generates the quantum superposition of coherent states, there is an extended explanation of this term at the Appendix \ref{app:QuadrupolarHamiltonian}). The third one,  the radio frequency term, is the interaction between the magnetic moment of the nucleus and a time-dependent weak magnetic field parallel to the $xy$-plane oscilating at the angular frequency $ \omega _{RF} $, denoted by $ \hbar  \gamma B_{1} \left( \hat{\mathbf{I}}_{x}\cos \left(\omega _{RF}t +  \upsilon \right) +\hat{\mathbf{I}}_{y}\sin \left(\omega _{RF}t +  \upsilon \right) \right)$, where   $\upsilon$ is the phase of the weak magnetic field $B_{1}$.  Those three terms constitute the time-independent Hamiltonian in Eq. (\ref{HamiltonianoRMN}) when represented in a rotating frame that rotates with frequency $\omega_{L}$ around the $z-$axis  
\begin{eqnarray}
\hat{\mathcal{H}}_{\text{NMR}}& = &-\hbar \left( \omega _{L}-\omega _{RF}\right) \hat{\mathbf{I}} _{z}+\hbar \frac{\omega _{Q}}{6}\left( 3\hat{\mathbf{I}}_{z}^{2}-\hat{\mathbf{I}}^{2} \right)   \notag \\
& &+\hbar \omega _{1}\left( \hat{\mathbf{I}}_{x}\cos \upsilon +\hat{\mathbf{I}}_{y}\sin \upsilon \right)   \text{,}    \label{HamiltonianoRMN}
\end{eqnarray}
where $ \omega _{L}=  \gamma B_{0}$ is the Larmor angular frequency and  $ \omega _{1}=  \gamma B_{1}$.

One of the main points of this development is to rewrite the NMR Hamiltonian of Eq. (\ref{HamiltonianoRMN}) into the effective Hamiltonian of Eq. (\ref{HamiltonianoEffetivo}). To do that, let us assume the radio frequency is tuned at $ \omega _{RF} =  \omega _{L}-p\omega _{Q}/2$ with $p\in\mathbb{Z}$, and its intensity is null, $ \omega _{1} = 0$. In this case, the NMR Hamiltonian can be written as
\begin{equation}
\hat{\mathcal{H}} =- \hbar \frac{\omega _{Q}}{2} \left(   p \hat{\mathbf{I}} _{z}-   \hat{\mathbf{I}}_{z}^{2}+ \frac{1}{3}\hat{\mathbf{I}}^{2}    \right)   \text{.}    \label{HamiltonianoNaoLinear}
\end{equation}
This nuclear spin Hamiltonian matches the effective Hamiltonian of Eq. (\ref{HamiltonianoEffetivo}) with the following physical parameters: the quadrupolar coupling $ \frac{\omega _{Q}}{2}$ and the coefficient $p$, which are equivalent to the effective coupling constant $\kappa$ and coefficient $ 2 \ \overline{n} + 1$, respectively. The Casimir operator represents an offset of energy generating a global phase on any quantum state. In this appropriate representation, the nuclear spin Hamiltonian of Eq. (\ref{HamiltonianoNaoLinear}) can generate Schr\"{o}dinger's cat states or, in other words, a superposition of coherent states.

Let us consider the definition of coherent state for a nuclear spin system $I$ as \cite{auccaise2013,perelomov1985Book}
\begin{equation}
\left\vert \zeta \left( \vartheta ,\varphi \right) \right\rangle =\sum_{m=I}^{-I}\frac{\zeta ^{I+m}}{\left( 1+\zeta ^{\ast }\zeta \right) ^{I} } \sqrt{\left(  
\begin{array}{c}
2I \\ 
I+m%
\end{array}%
\right) } \left\vert I,m\right\rangle \text{,} \label{DefinicaoEstadoCoerente}
\end{equation}
where  $\zeta =\tan \frac{ \vartheta }{2}\exp \left[ -i\varphi \right] $ represents the excitation parameter with angular values $0 \leq \vartheta \leq \pi$ and  $0 \leq \varphi \leq 2\pi$,  and $ \left( 
\begin{array}{c}
2I \\ 
I+m%
\end{array}%
\right) =\frac{\left( 2I\right) !}{\left( I+m\right) !\left( I-m\right) !}$ is the binomial coefficient.  The coherent state must be transformed by the non-linear unitary propagator and it is denoted by   $\left\vert \Psi \left( t\right) \right\rangle =\exp \left[- i\hat{\mathcal{H} }t/\hbar \right] \left\vert \zeta \left( \vartheta ,\varphi \right) \right\rangle $, where the non-linear unitary propagator transforms each element of the Dicke basis established by  $\left\vert \Psi_{m} \left( t\right) \right\rangle =\exp \left[- i\hat{\mathcal{H} } t / \hbar \right]  \left\vert I,m\right\rangle $. Therefore, as the Hamiltonian of Eq. (\ref{HamiltonianoNaoLinear}) describes the dynamics of the spin system, each element of the quantum basis is transformed as follows
\begin{equation}
\left\vert \Psi_{m} \left( t\right) \right\rangle = \exp \left[ i\frac{\omega _{Q}t}{2} \left(   p m-   m^{2}+ \frac{I\left(  I+1\right) }{3}    \right)  \right]  \left\vert I,m\right\rangle  \text{.} \label{CalculoPropagador}
\end{equation}
Next, using the condition $\frac{\omega _{Q}t}{2} = \frac{\pi}{2}$ and isolating the $t$ parameter, we define $t_{S}=\frac{\pi}{\omega_{Q}}$  as the time at which the Schr\"{o}dinger's cat state must be prepared and detected. Moreover, considering periodical boundary conditions and, simultaneously,  the non-linear unitary transformation to expand in a  Fourier series, as detailed on Eq. (18) and Eq. (19) of Ref. \cite{agarwal1997} for the even case, the initial coherent state defined in Eq. (\ref{DefinicaoEstadoCoerente}) is transformed into
\begin{widetext}
\begin{equation}
\left\vert \Psi \left( t_{S} \right) \right\rangle =\psi_{I} \left( \exp \left[ i \frac{\pi }{2}\left( p-1\right) I\right] \left\vert \zeta \left( \vartheta ,\varphi +\frac{\pi }{2}\left( p-1\right) \right) \right\rangle +\exp  \left[ i\frac{\pi }{2}\left( p+1\right) I\right] \left\vert \zeta \left( \vartheta ,\varphi +\frac{\pi }{2}\left( p+1\right) \right) \right\rangle \right) \text{,}  \label{CatStateGeneral}
\end{equation}
\end{widetext}
where $\psi_{I}  = \frac{1 }{\sqrt{2}} \exp  \left[ -i\frac{\pi }{2}\frac{2I\left( 2I+2\right)-3 }{12} \right]$ is a global phase.

The quantum state of Eq. (\ref{CatStateGeneral}) satisfies the theoretical predictions of the Ref. \cite{agarwal1997,klimov1998} at which the coefficient $ 2 \ \overline{n} + 1 = p$ must be positive, even reaching its limit value for the mean number of photons on the cavity $ \overline{n} = 0$, such that $p=1$, and the quantum superposition is 
\begin{equation}
\left\vert \Psi \left( t_{S} \right) \right\rangle =\psi _{I}\left( \left\vert \zeta \left( \vartheta ,\varphi \right) \right\rangle +\exp \left[ i\pi I\right] \left\vert \zeta \left( \vartheta ,\varphi +\pi \right) \right\rangle \right) \text{.} \label{SuperpositionAtomLight}
\end{equation}
On the other hand, in the quadrupolar spin system, the parameter $p$ could be positive, negative, and null. The boundary case is when $p$ is null, it means that the quadrupolar system evolves freely at the resonant frequency, and the quantum superposition is 
\begin{equation}
\left\vert \Psi \left( t_{S} \right) \right\rangle =\psi _{I}^{\prime} \left( \left\vert \zeta \left( \vartheta ,\varphi_{+}\right) \right\rangle +\exp \left[ -i\pi I\right] \left\vert \zeta \left( \vartheta ,\varphi_{-}\right) \right\rangle \right) \text{,} \label{SuperpositionNuclearSpin}
\end{equation}
where $ \psi _{I}^{\prime} = \psi _{I} \exp \left[ i\frac{\pi }{2}I\right]$ and $  \varphi_{\pm}=\varphi \pm \frac{\pi }{2} $. This kind of quantum superposition  mimics a counter-intuitive physical interpretation at the atom-field scenario with  $ \overline{n}= \frac{p-1}{2}$ at $p \leq 0$ representing negative mean values of photons in the cavity. In NMR spin systems, the off and on-resonance conditions of the spectrometer are degrees of freedom achieved from the experimental setup, which is different from the atom-field scenario established by the initial quantum state of the reservoir.

One of the main characteristics of this nuclear spin scenario is the quantum control on the choice of the type of superposition to be implemented. For even values of $p$, the quantum state  denoted by Eq. (\ref{SuperpositionNuclearSpin}) represents the superposition of two coherent states, which are orthogonal to the initial one. On the other hand, for odd values of $p$, the quantum state denoted by Eq. (\ref{SuperpositionAtomLight}) represents the superposition of two coherent states parallel to the initial one.

Also, in this study is performed an interesting theoretical and mathematical procedure to identify graphically on a phase space the degree of quantum correlations of the  coherent state superpositions; it is the application of the Wigner quasiprobability distribution function definition \cite{agarwal1981,benedict1999,sanchez-soto2013}, which was applied in optics  \cite{ourjoumtsev2006,ourjoumtsev2007}, atoms  \cite{monroe1996,leibfried1996}, nuclear spins \cite{auccaise2013,auccaise2015,teles2015}, and described theoretically for spin systems \cite{garon2015,koczor2019JPA,koczor2019AP}. Accordingly with these references, and  for any density matrix denoted by  $ \hat{\boldsymbol{\rho }} $, the definition obeys the mathematical expression
\begin{equation}
W\left(  \theta ,\phi \right)  = \sqrt{\frac{2I+1}{4\pi}}
{\displaystyle\sum\limits_{K=0}^{2I}} \ {\displaystyle\sum\limits_{Q=-K}^{K}}
 \  T_{KQ} \  Y_{KQ}\left(  \theta,\phi\right)
\text{,} \label{WignerDistributionFunction}
\end{equation}
where angular parameter values $\theta \in \left[ 0, \pi \right]$ and $\phi \in \left[ 0, 2\pi \right]$,    $T_{KQ} = \mathtt{Tr} \left\{ \hat{\boldsymbol{\rho }} \ \hat{\mathbf{T}}_{KQ}^{\dagger} \right\} $ denotes the operator $\hat{\mathbf{T}}_{KQ}$ mean value, $\hat{\mathbf{T}}_{KQ} $  and $Y_{KQ}\left( \theta ,\phi \right) $ denote the spherical tensor operator and the spherical harmonic function  of rank $K$ and order $Q$, respectively.

\section{Description of experimental procedures}
\label{sec:DescriptionExperimentalProcedures}

\begin{figure*}[!ht]
\includegraphics[width=1.00\textwidth]{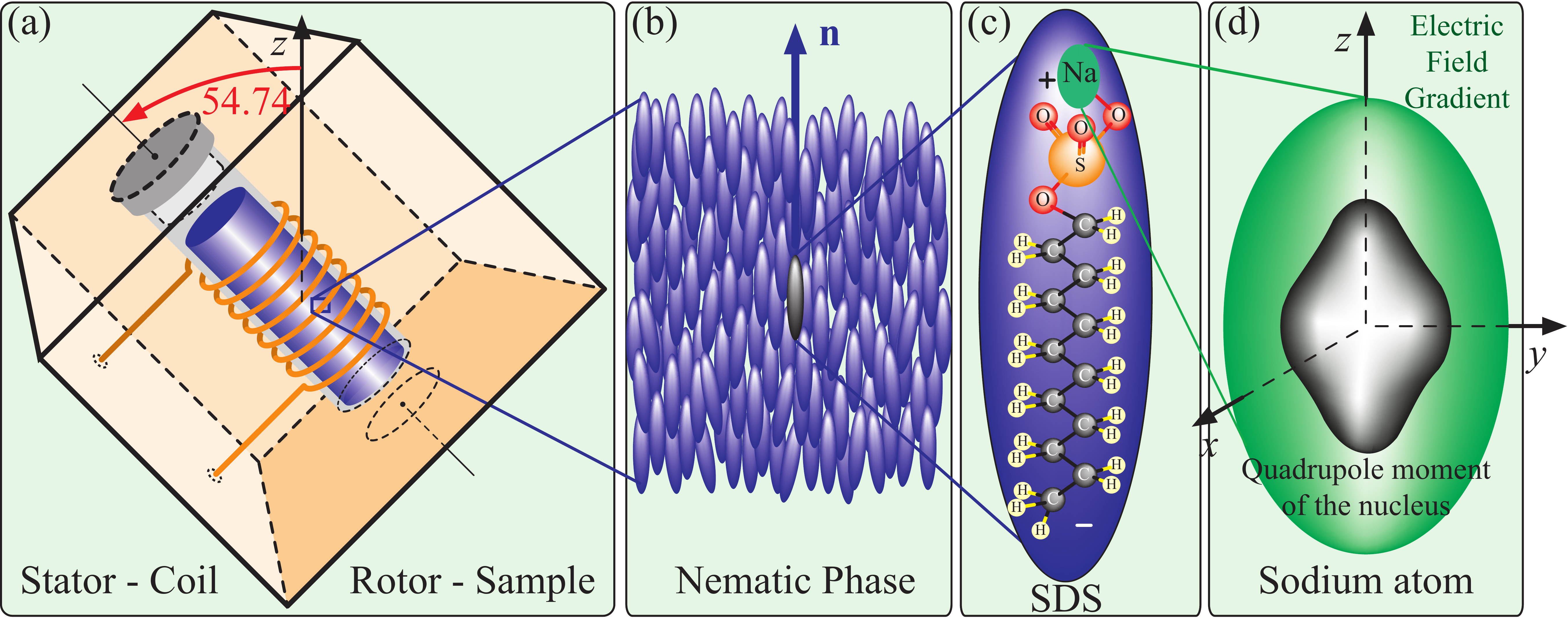}
\caption{(Color online) Pictorial representation of some characteristics of the experiment and sample. (a) The NMR probe consists of a spinning system composed of a stator, a rotor, where the sample is inserted, and a rf coil, which surrounds the rotor. The sample is oriented along the characteristic magic-angle value of 54.74  degrees, however, despite using a magic-angle spinning probe, the sample is kept static. This probe was used only due to the hard rf pulses it can provide. (b) The nematic phase features a long-range orientational order of  elongated molecules that can freely rotate about all their axes, but point on average in the same direction characterized by the director $\textbf{n}$, which depicts the local preferable direction of molecules within a certain volume. (c) Sodium Dodecyl Sulfate (SDS) with molecular formula Na-C$_{12}$H$_{25}$SO$_{4}$, the plus and minus signs stablish the polar bonds due to its electronegativity between the bonded atoms. (d) The green ellipsoid depicts the electric field gradient around the nucleus generated by charged particles of the molecule and the black central region depicts the quadrupole moment of the nucleus \cite{slichter1992Book}. } \label{fig:ExperimentalDescription}
\end{figure*}

The experimental setup around implementing the Schr\"{o}dinger's cat state was achieved using a Tecmag Discovery Console, a Jastec 9.4 T superconductor magnet and a Jakobsen 5 mm solid-state NMR probe.  A lyotropic liquid crystal sample was placed into a 4 mm o.d. zirconia (ZrO$_{2}$) rotor and sealed with a Kel-F cap (see Fig. \ref{fig:ExperimentalDescription}(a)). One of the most interesting properties of these lyotropic liquid crystals is the collective orientation capability to achieve appropriate arrangements. From those arrangements, the nematic phase is highlighted by its typical quality of a long-range orientational order of elongated molecules pointing on average in the same direction established by the director   $\textbf{n}$ (see Fig. \ref{fig:ExperimentalDescription}(b)). In this sense, the samples used in this experimental development are classified at the nematic phase   by their stoichiometric composition and physical properties, as detailed in Ref. \cite{auccaise2008,quist1992} for the Sodium Dodecyl Sulfate (SDS) sample. A pictorial scheme about the atomic composition for the SDS sample is depicted in Fig.  \ref{fig:ExperimentalDescription}(c)  with molecular formula Na-C$_{12}$H$_{25}$SO$_{4}$.

\textit{Sodium nuclei.-} The  $^{23}$Na   nucleus (100 \% abundant in nature) has spin value $I=3/2$, which allows establishing the dimension of the Hilbert space $d=2I+1=4$.   The spectrometer operates at the Larmor frequency of  $\omega_{L}/2\pi =   105.571$ MHz. The quadrupolar coupling measured from the satellite lines is $15220 \pm 70$Hz. The experiment was performed using spectral widths of 41666 Hz. The sample temperature was fixed at 28$^{\circ}$C. The $\frac{\pi}{2}$-pulse was calibrated at 10 $\mu$s. Acquisition time was 49.152 ms{, the number of points was 4096, and the dwell time was 12 $\mu$s}. The recycle delay was 250 ms.  The longitudinal and  transverse relaxation times were measured $T_{1}\approx 20  $ ms and  $T_{2}\approx 1.46  $ ms, respectively. The lyotropic liquid crystal at the nematic phase related with the  SDS sample was prepared  at stoichiometry values denoted by $21.28$ \%  of SDS, $3.56$ \%  of decanol and $75.16$ \%  of deuterium oxide.

\subsection{Initialization of the quantum states}

\textit{Initialization.-} Standard high temperature NMR description of the quantum state is expressed as a first order expansion of the density matrix definition \cite{oliveira2007Book}
\[
\hat{\boldsymbol{\rho }}=\frac{1}{\mathcal{Z}}\exp \left[ -\beta \hat{ \mathcal{H}}_{0}\right] \simeq \frac{1}{\mathcal{Z}}\hat{\mathbf{1}} _ { d \times d }-\frac{\beta}{\mathcal{Z}} \hat{\mathcal{H}}_{0}\text{,}
\]
where $\mathcal{Z}$ is the partition function, $\beta =\left( k_{B}T\right) ^{-1}$, $k_{B}$ is the Boltzmann constant and $T$ the room temperature,  $ \hat{\mathcal{H}}_{0}=- \hbar \omega _{L}\hat{\mathbf{I}}_{z}  $   is the Zeeman Hamiltonian at the laboratory frame so that the density matrix is expressed as
\begin{equation}
\hat{\boldsymbol{\rho }}\simeq \frac{1}{\mathcal{Z}}\hat{\mathbf{1}} _ { d \times d }   + \frac{ \beta \hbar \omega _{L} }{\mathcal{Z}} \hat{\mathbf{I}}_{z}   \text{.}
\end{equation}
The main purpose of the initialization procedure is to transform the second term of the expanded density matrix into a contribution with equivalent properties of an effective pure state. In order to do that, we use the temporal average procedure \cite{fortunato2002,teles2007}. Then, the average density matrix is represented as
\begin{equation}
\hat{\boldsymbol{\rho }}\simeq \left( \frac{1}{\mathcal{Z}} - \epsilon \right) \hat{ \mathbf{1} } _ { d \times d } + \epsilon \Delta\hat{\boldsymbol{\rho }} \text{,}
\end{equation}
where $\epsilon =\frac{ \beta \hbar \omega _{L} }{ \mathcal{Z}}   $ is approximately  $ 0.426\times 10^{-5}$ for $^{23}$Na nuclei   and  $\Delta\hat{\boldsymbol{\rho }} \equiv\left\vert \zeta\left(  \vartheta,\varphi\right)  \right\rangle \left\langle \zeta\left(  \vartheta,\varphi\right)  \right\vert $ is the deviation density operator \cite{auccaise2013}. An appropriate angular parameters choice, $\vartheta=\pi/2$ and $\varphi=0$, implies in the generation of the initial coherent state $ \left\vert \zeta\left(  \pi/2,0\right) \right\rangle $.

\begin{figure}[!ht]
\includegraphics[width=0.48\textwidth]{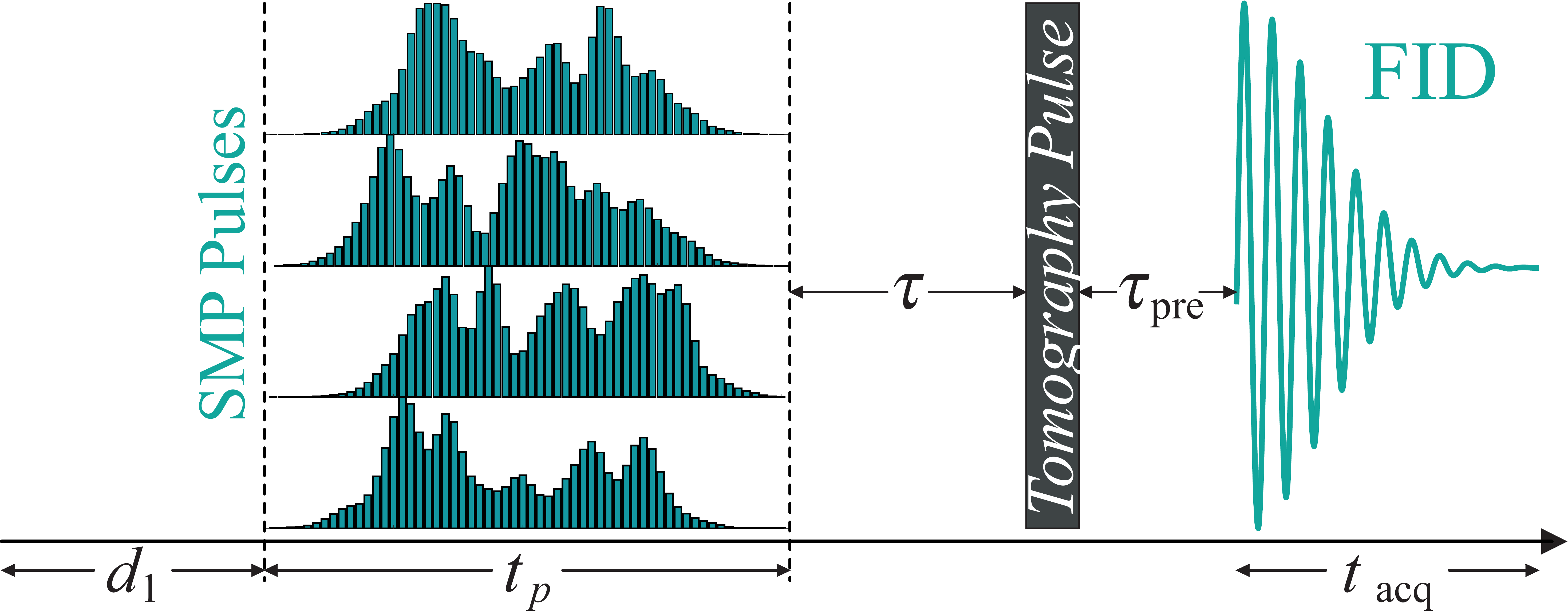}
\caption{(Color online)  The NMR Pulse sequence starts with a recycle time $d_{1}$, then, the strongly modulated pulse is applied during the time $t_{p}$, allowing the preparation of the initial quantum state  $\left\vert  \psi_{0}  \right\rangle \equiv \left\vert \zeta\left(  \pi/2, 0 \right) \right\rangle $. The generation of the cat state is performed by allowing the free evolution of the spin system during the delay time $\tau = 1/\left( 2 \nu_{Q}\right)$. Then the tomography procedure is implemented using one hard pulse, a delay time of pre-acquition $\tau_{\text{pre}} = 1/  \nu_{Q} $ to protect the receiver, and turning on the detector during $t_{\text{acq}}$ to read out the free induction decay (FID). } \label{fig:PulseSequence}
\end{figure}

\textit{Strongly Modulating Pulses (SMP).-}  The technique of SMP  was adapted in order to smooth the modulation of the radio frequency pulse \cite{fortunato2002}. The technique is an optimization procedure where  the  time $t_{p}$ of the modulated pulse is divided into $n$ temporal slices, and each interval time is fixed at $\delta t$, such that $t_{p} = n \delta t$. At each interval, labelled  with the subscript $k=1,\ldots,n$, the strength and phase of the pulse, $\omega_{k}$ and $\phi_{k}$, must be defined, which allows determining a set of physical parameters that depicts one modulation.  If more than one modulation is considered then it configures the temporal average procedure \cite{fortunato2002,teles2007,araujo-ferreira2013,auccaise2015,auccaise2013}. In this experimental implementation, four strongly modulating pulses were used, as pictorially sketched in Fig. \ref{fig:PulseSequence}, where four sets of  $\omega_{k}$ are displayed. At the end of this stage the initial quantum state   $\left\vert  \psi_{0}  \right\rangle \equiv \left\vert \zeta\left(  \pi/2, 0 \right) \right\rangle $ must be prepared.

\subsection{Quantum state tomography procedure}
\label{sec:QuantumStateTomographyProcedure}

The quantum state tomography procedure is the method employed to reconstruct the density matrix. In the case of this experimental implementation, the technique of global rotations was used \cite{teles2007}. In this technique, the representation of Euler angles is used to map the effect of a unitary rotation performed by the tomography pulse (see Fig. \ref{fig:PulseSequence}) and an appropriate phase cycling. From the representation of Euler angles, an optimal rotation is identified that maximizes the best signal strength when related to the angular parameter $\theta_{qst}$ of the reduced Wigner function $d^{L}_{m^{\prime},m}\left(\theta_{qst} \right)$ with $L=0,1,\ldots,2I$ and $m^{\prime},m=-L,\dots,+L$ (see Eq. (2) of Ref. \cite{varshalovich1988Book}). Furthermore, the proper choice of the subscripts $L$ and $m$ allows writing any density matrix as a linear combination of irreducible tensors as follows
\begin{equation}
\boldsymbol{\hat{\rho}} =\sum\limits_{L=0}^{2I}\sum_{m=-L}^{+L}a_{L,m}\hat{\mathbf{T}}_{L,m}\text{.} \label{Rho-IrreducibleTensor}
\end{equation}
From the phase cycling, one pre-established coherence order of the density matrix is selected, eliminating the contribution of the unwanted ones. For example, in the case of a spin $I=3/2$,  to detect the zero order coherences the angular parameter related to the nutation angle is set to $\theta_{qst} = \frac{\pi}{2}$, and the angular parameters $\phi_{qst} =\left\{ \frac{\pi }{2},\pi , \frac{3\pi } { 2} ,0 \right\} $ and $\alpha_{qst} =\left\{ 0,\frac{3\pi }{2},\pi ,\frac{\pi }{2} \right\} $, which configure the phases of the tomography pulse (transmitter)  and the phase of the receiver, respectively. These values are listed on Tab. I and Tab. II of Ref. \cite{teles2007}, along with other angular parameter values needed to detect other coherence orders.

\begin{figure}
\includegraphics[width=0.49\textwidth]{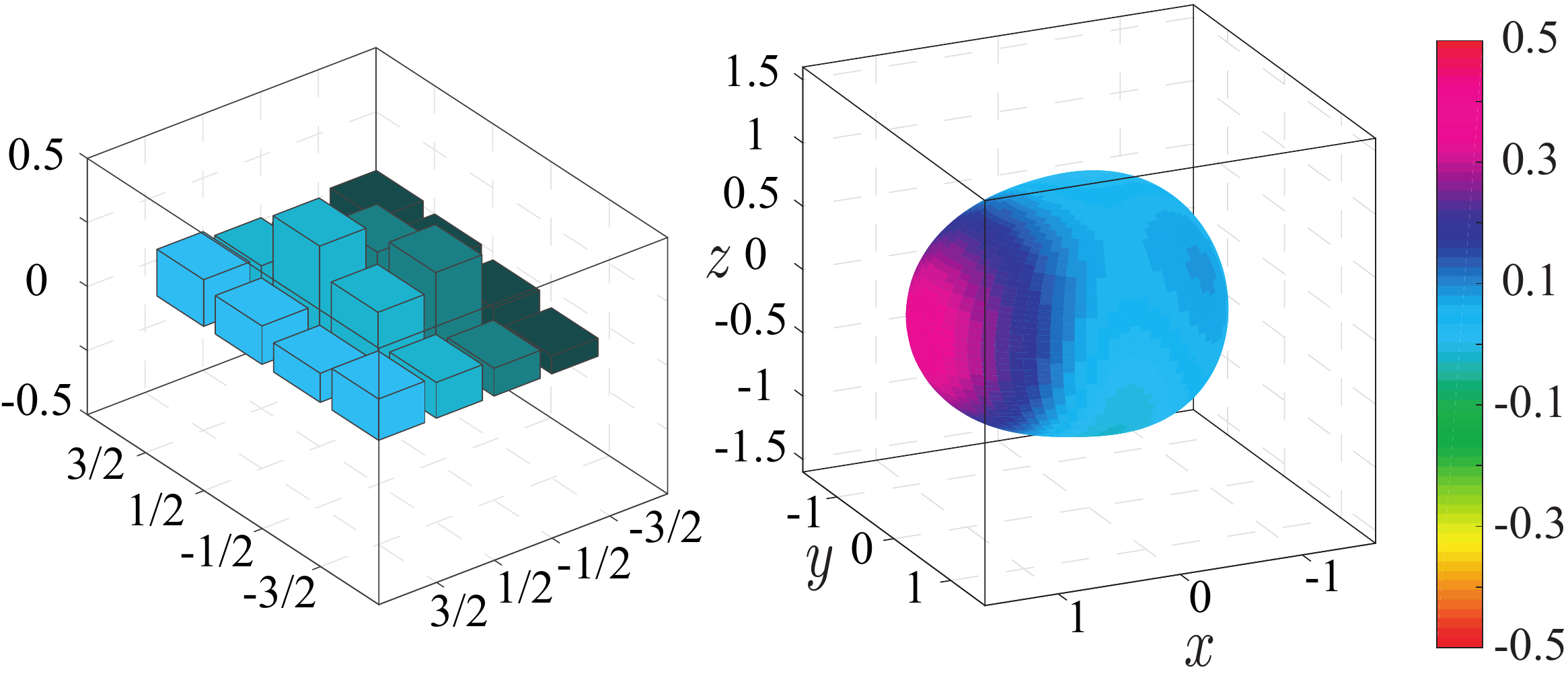}
\caption{(Color online) On the left side, the experimental density matrix real elements of the initial coherent state   $\left\vert \zeta\left(\pi/2,0 \right)\right\rangle$ are represented by a 2D bar chart, with Fidelity value of 0.986. On the right side, the Wigner quasi-probability  distribution function is shown as a surface in a three dimensional angular momentum frame denoted by $k=x,y,z$. The surface is generated using the experimental density matrix of the initial coherent state   $\left\vert \zeta\left(\pi/2,0 \right)\right\rangle$. As expected, the highest distribution probability points along the positive $x$-axis of the frame.  The   colour bar encodes the quasiprobability distribution function values. }  \label{fig:Preparado3o2}
\end{figure}

The experimental implementation is summarized in the following stages: \textit{(1$^{\circ}$) } executing the modulation of the radio frequency pulse to initialize the quantum state  $\left\vert  \psi_{0}  \right\rangle \equiv \left\vert \zeta\left(  \pi/2, 0 \right) \right\rangle $,  \textit{(2$^{\circ}$) }  performing the tomography procedure using the angular parameters values for $\theta_{qst}$,  $\phi_{qst}$ and $\alpha_{qst}$,  \textit{(3$^{\circ}$) }  acquiring the NMR signal and analyzing its respective spectrum.  

Information about the elements of the density matrix is obtained through the intensity of the spectral lines. The  theoretical procedure to relate the intensities of the spectral lines to the elements of the density matrix follows a linear system of equations that can be represented by the standard algebraic equation:
\begin{equation}
AX =  B\text{,} \label{LinearSystem}
\end{equation}
where $A$ is any matrix with numerical values computed using the mathematical procedures of quantum mechanics to describe the unitary rotation and the detection of the free induction decay, $X$ is the vector column with elements related to the coefficients $a_{L,m} $ of the linear combination of irreducible tensor operators to represent the density matrix of Eq. (\ref{Rho-IrreducibleTensor}), and $B$ is the vector column with element values representing the line intensities of each spectrum. For example, solving this linear system of equations makes it possible to find each element of the density matrix of the quantum state  $\left\vert \zeta\left(\pi/2,0 \right)\right\rangle$   which is displayed as a bar chart at the left side of Fig. \ref{fig:Preparado3o2}. The quality of the implemented quantum state is quantified using the fidelity definition of the Eq. (3) in Ref. \cite{fortunato2002}, such that the computed value is 0.986 or 98.6 \% of similarity when compared with the theoretical density matrix. 

On the other hand, a parallel procedure to test the efficiency of the initial coherent quantum state is the application of the Wigner quasiprobability distribution function for the tomographed quantum state. The result is shown at the right side of Fig. \ref{fig:Preparado3o2}. One of the main characteristics of the surface representation is the spatial orientation of the highest probability values, which is along the positive $x$-axis, as expected, and by this preferred orientation it is known as an $x$\textit{-coherent quantum state}  \cite{jin2007PRA,auccaise2015}. 

The previous paragraphs show the existence of imperfections about this experimental implementation, and in that  manner, three primary sources of error can be pointed out. The first one is due to the  room temperature fluctuations despite the electronic sample temperature control. One of the well known properties of the quadrupole moment is the dependence on the temperature. The temperature fluctuation induces a fluctuation on the quadrupolar coupling of the nucleus which is identified on the broadening of the spectral lines. This broadening effect introduces an error of $\pm 70$Hz  on the quadrupolar coupling  as detailed on the experimental parameters of the topic ``Sodium nuclei'' of this section. Another source of error is due to hidden delays before and after the application of any radio-frequency pulse. The turn on-off  of any  electronic device is not instantaneous and the time needed to achieve its designated power level is denominated as a transient time, which for standard NMR spectrometers can be identified  at  $\sim 2-3 \mu$s. These time delays will generate undesired evolutions of the quantum system. Another source of error is due to imperfections on the modulation of the shaped pulse. The response of the electronic device into the generation of the accurate amplitude and phase modulation and their application along the sample can suffer slight modifications. Therefore, the collective action of these three source of errors  will be  encoded on the tomographed quantum state and their primary consequence is diminishing the fidelity value.

\section{Experimental   results}
\label{sec:ExperimentalResults}

\begin{figure*}
\includegraphics[width=0.99\textwidth]{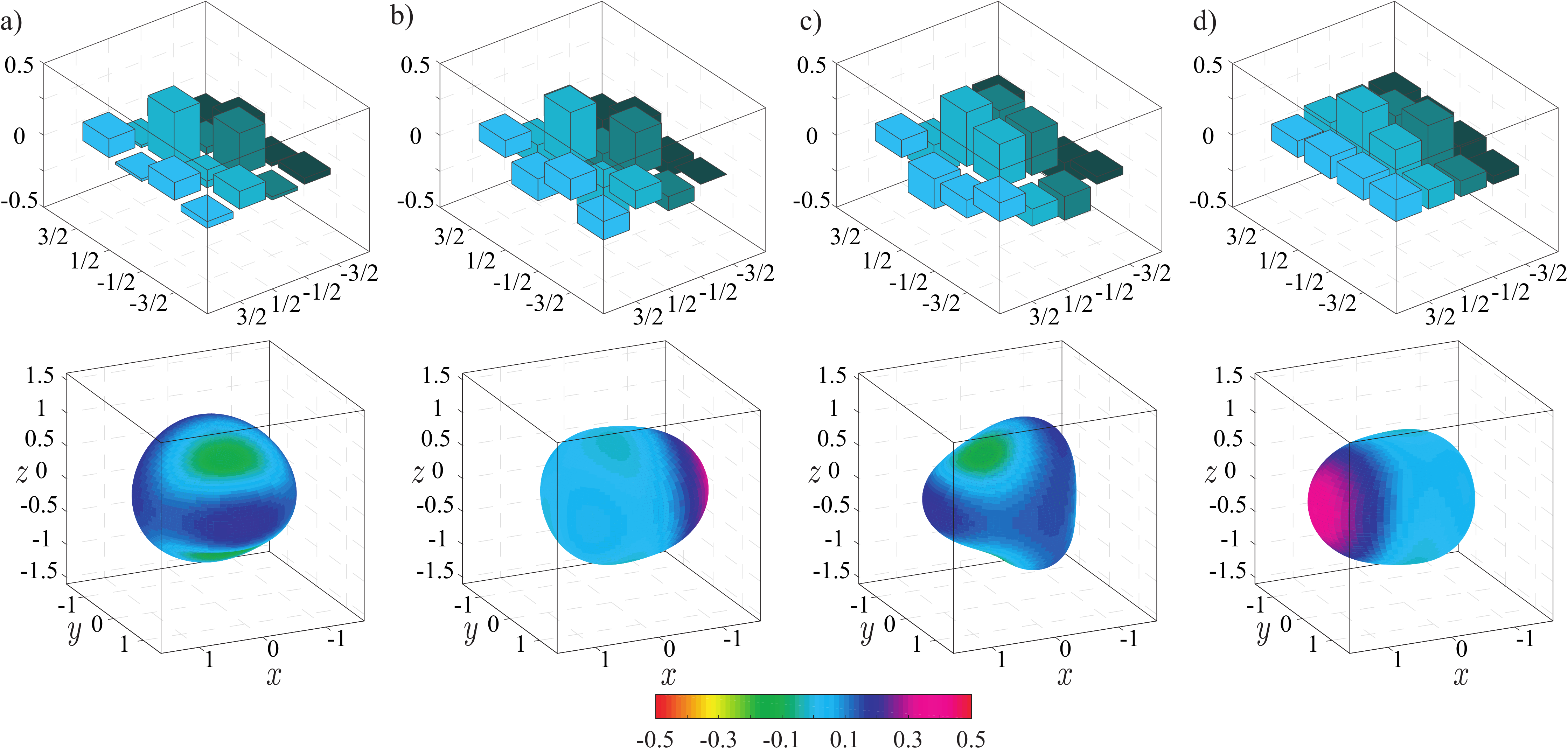}
\caption{(Color online) 
On the top are shown the  experimental deviation density matrix real elements  as bar charts (imaginary elements were computed but not shown) generated by the quantum state tomography procedure \cite{teles2007}. The Wigner quasiprobability distribution function is shown on the bottom as a surface plot  in a three-dimensional angular momentum frame labeled by $x,y,z$ \cite{agarwal1981,benedict1999,sanchez-soto2013,auccaise2015}. (a)  The Schr\"{o}dinger's cat state at $p=1$ representing Eq. (\ref{SuperpositionAtomLight}) at the time $t_{S}=1/\left(2\nu_{Q}\right) = 32.85  \mu \texttt{s} $ with a fidelity value of $0.980$. (b) The coherent state  $\left\vert \zeta\left(\pi/2,\pi \right)\right\rangle$ at $p=1$ at the time $2t_{S}=  1/ \nu_{Q}  = 65.70  \mu \texttt{s} $ with a fidelity value of  $F=0.969$. (c)  The Schr\"{o}dinger's cat state at $p=0$ representing the Eq. (\ref{SuperpositionNuclearSpin}) at the time $t_{S}=  1/\left(2\nu_{Q}\right) = 32.85  \mu \texttt{s} $ with a fidelity value of $0.970$. (d) The coherent state  $\left\vert \zeta\left(\pi/2, 0 \right)\right\rangle$ at $p=0$ generated under the free evolution at time    $2t_{S}=  1/ \nu_{Q} = 65.70  \mu \texttt{s} $ with the fidelity of $F=0.960$.  The   color bar under the boxes encodes the quasiprobability distribution function values. }
\label{fig:Resultado3o2}
\end{figure*}

The generation of the Schr\"{o}dinger's cat state is achieved performing a free evolution dynamics of a quadrupolar spin system characterized theoretically by the effective Hamiltonian of Eq. (\ref{HamiltonianoNaoLinear}), at  the time $t_{S}=32.85 \mu$s  and under two values of the $p$ parameter: $p=1$ and $p=0$. In the first case, $p=1$, monitoring the spin system, the superposition of coherent states of Eq. (\ref{SuperpositionAtomLight}) can be detected  at time $t_{S}=\frac{1}{2\nu_{Q}} = 32.85 \mu \texttt{s} $, and the experimental density matrix real elements, $Re \lbrace \Delta\hat{\boldsymbol{\rho }}  \left( t_{S} \right) \rbrace$, are depicted as bar charts in Fig. \ref{fig:Resultado3o2}(a). Going twice this characteristic time and continuing monitoring the spin system, it evolves to the coherent state with an opposite initial $\varphi$-phase denoted by   $\left\vert \zeta\left(\pi/2,\pi \right)\right\rangle$ at time $t=2t_{S}=\frac{1}{\nu_{Q}} = 65.7  \mu \texttt{s} $ and the  experimental density matrix real elements, $Re \lbrace \Delta\hat{\boldsymbol{\rho }}  \left(2 t_{S} \right) \rbrace \equiv  Re \lbrace\left\vert \zeta\left(\pi/2,\pi \right)\right\rangle    \left\langle \zeta\left(\pi/2,\pi \right) \right\vert  \rbrace $, are shown in Fig. \ref{fig:Resultado3o2}(b). If the evolution time  assumes $3t_{S}$ and $4t_{S}$ then the superposition of Eq. (\ref{SuperpositionAtomLight})  and the initial coherent state are recovered, respectively, completing a cycle of the dynamics.  In the second case, $p=0$, monitoring the spin system, the superposition of coherent states of Eq. (\ref{SuperpositionNuclearSpin}) can be detected  at time $t_{S}=\frac{1}{2\nu_{Q}} = 32.85 \mu \texttt{s} $, and the experimental density matrix real elements, $Re \lbrace \Delta\hat{\boldsymbol{\rho }}  \left( t_{S} \right) \rbrace$, are depicted as bar charts in Fig. \ref{fig:Resultado3o2}(c). Going twice this characteristic time, the spin system evolves at the coherent state with the same initial  $\varphi$-phase denoted by   $\left\vert \zeta\left(\pi/2,0  \right)\right\rangle$ at time $t = 2t_{S}=\frac{1}{\nu_{Q}} = 65.7  \mu \texttt{s} $ and the experimental density matrix real elements, $Re \lbrace \Delta\hat{\boldsymbol{\rho }}  \left(2 t_{S} \right) \rbrace \equiv  Re \lbrace\left\vert \zeta\left(\pi/2,0 \right)\right\rangle    \left\langle \zeta\left(\pi/2,0 \right) \right\vert  \rbrace $, are shown in Fig. \ref{fig:Resultado3o2}(d). Therefore, the system come back to the initial coherent state, completing a cycle of the dynamics.

Additionally, the application of the Wigner quasiprobability distribution function definition for the tomographed deviation density matrices on the top of Fig. \ref{fig:Resultado3o2}  allow generating surfaces at the phase space as shown at the bottom of Fig. \ref{fig:Resultado3o2}. The color bar encodes the intensity value of the distribution function. We found that this representation has the advantage of the interference pattern which is characteristic of quantum states with non-classical correlations as it is observed as a kind of  wave closed surface   between positive and negative values of the distribution function intensities. Basically, there are $2I$  green deep regions interleaved by other $2I$ blue raised regions, which characterizes the interference pattern.

\section{Discussions}
\label{sec:Discussions}

Since the fundamentals of Quantum Mechanics were proposed, the dynamics of particles at microscopic scales had just emerged from a new point of view. The understanding processes of the quantum theory had some emblematic phenomena that challenged the standard interpretation at that time, such as   Schr\"{o}dinger's cat states. In that sense, the main purpose of the generation of the Schr\"{o}dinger's cat state  using quadrupolar spin systems is the production of non-classical correlations taking advantage of the internal degrees of freedom about the spin description of a nucleus.    These correlations are generated by an appropriate superposition of the Dicke basis elements $ \left\vert I,m\right\rangle$, which are transformed under the action of  the propagator   $\exp \left[ -i\hat{\mathcal{H} }t_{S}/\hbar \right]$, such that each of them gains a phase as analogously happens in the collective behavior of many atoms to produce a type of superposition of coherent states  \cite{neergaard-nielsen2006, ourjoumtsev2007, vlastakis2013}. From it, quantum states of quadrupolar spin systems achieve optimal metrological properties used to reach Heisenberg-limited sensitivity  \citep{lucke2014}.

The dimensionality of the Hilbert space is an important point to be highlighted. In this sense, it is known that a quantum tomography procedure is a hard task to be implemented in quantum systems with a large number of particles \cite{lucke2014}, which are described in  Hilbert spaces of higher dimensions. On the other hand, quantum systems with low dimensional Hilbert spaces as in quadrupolar spin system favours the implementation of the quantum state tomography procedure to characterise and to monitor its quantum state, as happens routinely for few particles (spins \cite{teles2007}, photons \cite{leibfried1996}), which could be used as a test workbench of the principles of quantum theory.
 
The source of the generation of cat states using quadrupolar nuclei spin systems is the quadrupolar coupling. The quadrupolar coupling represents the average effects of the interaction between any electric field gradient around the nucleus and the quadrupole moment of the nucleus \cite{slichter1992Book}. From a theoretical point of view, every nucleus with $I>1/2$ and $\omega_{Q} \neq 0$ can be used as a quantum computer to implement  quantum simulations, or any quantum information task. However, from an experimental point of view, there are other requirements to do that, such as the ability to access information about the system's quantum state. This is achieved on arrangements of spin systems with a spatial order, as happens in solid crystals  \cite{kampermann2005,teles2015,nie2015,teles2018}, at different semiconductor platforms  \cite{yusa2005,miranowicz2015,hendrickx2021,glenn2018},  nitrogen vacancies reported by many groups \cite{dutt2007,kong2016,rose2018,aslam2017,evans2018,lesik2019}, and liquid crystals \cite{das2003}.

The  Schr\"{o}dinger's cat states implemented in this report are a little different from the GHZ type Schr\"{o}dinger's cat states. The main difference is established in the kind of superposition of Dicke basis elements, such as denoted by the quantum state of Eq. (\ref{CalculoPropagador}) and Eq. (\ref{CatStateGeneral}). From these equations and for each Dicke basis elements, each probability amplitude have no-null value (see bar charts on the top of Fig. \ref{fig:Resultado3o2}) and depends on appropriate linear combinations of the binomial coefficient as defined in Eq. (\ref{DefinicaoEstadoCoerente}). For the GHZ type and considering angular momentum notation, there are two elements of the Dicke basis, which corresponds with no-null probability amplitude (see bar charts on Fig. 3 of Ref. \cite{song2017PRL})  and generally the elements  $ \left\vert J,+J\right\rangle$ and  $ \left\vert J,-J\right\rangle$ in the angular momentum notation or $ \left\vert 0,N\right\rangle \equiv  \left\vert 0\right\rangle_{1}  \left\vert 0\right\rangle_{2}\cdots  \left\vert 0\right\rangle_{N}    $ and $ \left\vert 1,N\right\rangle \equiv  \left\vert 1\right\rangle_{1}  \left\vert 1\right\rangle_{2} \cdots  \left\vert 1\right\rangle_{N}   $ in the qubit notation, where they were explored entangling atom-field Schr\"{o}dinger's cat states \cite{hacker2019}, Rydberg atom arrays \cite{omran2019}.

The versatility of the analysis we are introducing can be extended to other crystal system, for example in Ref. \cite{nie2015} the quantum state of the spin system was initialized in the quantum state $\left\vert +\frac{3}{2}\right\rangle $ (or in the papers notation $\left\vert 3,0\right\rangle $) and it will be transformed using a propagator with the Hamiltonian of Eq. (11) assuming $E_{c}=0$ of the Ref. \cite{gati2007} the generation of the Schr\"{o}dinger's cat state will be implemented with the interference pattern around the $x-y$ or $x-z$ plane.

\section{Conclusions}
\label{sec:Conclusions}
In summary, we were successful in carrying out $^{23}$Na NMR experiments dedicated to the generation of Schr\"{o}dinger's cat states, using a lyotropic liquid crystal in its nematic phase. The versatility of  the NMR setup depends on the parameter $p$, which allows achieving the atom-field approach for $p \geq 1$ and NMR approach for any null or negative value of $p$. The efficiency and accuracy of this study is verified performing the quantum state tomography procedure, the Wigner quasiprobability distribution  function definition, and fidelity values higher than 96\%.

The development implemented using this soft matter quantum approach highlights the strong and straight bond between solid-state quantum devices and an atomic quantum computer prototype. In addition, this approach is a kind of proof of principle showing that both quantum techniques will work well together  in the pursuit of a common purpose.

\section{Acknowledgements}

The authors  acknowledge  the National Institute of Science and Technology for Quantum Information (INCT-QI). A.C.S.L. acknowledges CNPq (142118/2018-4). E.L.O.  acknowledges CNPq (140215/2015-8).  T.J.B. acknowledges financial support from CNPq (308076/2018-4) and FAPESP (2012/02208-5). R.A. acknowledges CNPq (309023/2014-9, 459134/2014-0). This study was financed in part by the Coordena\c{c}\~{a}o de Aperfei\c{c}oamento de Pessoal de N\'{i}vel Superior - Brasil (CAPES) - Finance Code 001.


\appendix

\section{Theoretical Procedures}

\subsection{The quadrupolar Hamiltonian}
\label{app:QuadrupolarHamiltonian}

The quadrupolar Hamiltonian is the generator of the cat state. In this appendix, more details and explanations about the origin of this important quantum state are presented. The nature of the quadrupolar coupling has an electrical source \cite{slichter1992Book,wasylishen2012Book}, because it arises from the interaction of any distribution of positive charges on the atomic nuclei (the black non regular geometric volume  in Fig. \ref{fig:ExperimentalDescription}(d)) with an effective electric field gradient generated by the charges of the molecule itself or its neighbors (the green ellipsoid  in Fig. \ref{fig:ExperimentalDescription}(d)). The best formalism to explain this type of interaction was detailed in Chapter 10 of Ref. \cite{slichter1992Book} and here is presented the main aspects of this theory.

From the fundamentals of the electromagnetism, the electric potential energy of any charge distribution $ \varrho \left( \mathbf{r}\right)$ submitted to an electric potential $ V\left( \mathbf{r}\right)$ is  denoted by
\begin{equation}
E=\int \varrho \left( \mathbf{r}\right) V\left( \mathbf{r}\right) d^{3} \mathbf{r}\text{.}
\end{equation}
As a matter of fact, the phenomenon is reduced to a small spatial arrangement of particles (the nucleus), it enables applying the  Taylor's series definition and rewriting the electric potential around the origin of any  coordinate system  up to second order expansion and neglecting the higher ones. From this expansion  emerges the physical interpretation of each term, such that  the zero order term represents an energy offset; the first order term represents the energy of the electrical dipole moment of the nucleus such that at the stationary equilibrium the average  electric field around the nucleus is null; and  the second order term defines the quadrupolar energy contribution as denoted by
\begin{equation}
E_{Q} = \sum_{x_{j},x_{k}=x,y,z}\left. \frac{\partial
^{2}\left( V\left( \mathbf{r}\right) \right) }{\partial x_{j}\partial x_{k}}%
\right\vert _{\mathbf{r}=\mathbf{0}}\int x_{j}x_{k}\varrho \left( \mathbf{r}%
\right) d^{3}\mathbf{r}\text{.} \label{Expansion2ndTerm}
\end{equation}
This energy contribution is described in its operator representation applying the quantum mechanical notation of operators, the irreducible tensor operators, the Clebsch-Gordan coefficients and the Wigner-Eckart theorem such that the quadrupolar energy contribution of the Eq. (\ref{Expansion2ndTerm}) is represented by 
\begin{equation*}
\hat{ \mathcal{H}}_{Q} = \sum_{x_{j},x_{k}}\frac{eQV_{x_{j},x_{k}}}{6I\left( 2I-1\right) }%
\left( \frac{3}{2}\left( \hat{\mathbf{I}}_{x_{j}}\hat{\mathbf{I}}_{x_{k}}+%
\hat{\mathbf{I}}_{x_{k}}\hat{\mathbf{I}}_{x_{j}}\right) -\delta
_{x_{j},x_{k}}\hat{\mathbf{I}}^{2}\right) \text{,}
\end{equation*}
where $e$ is the elemental charge, $Q$ is the quadrupole moment of the nucleus,   $V_{x_{j},x_{k}}=\left. \frac{\partial ^{2}\left( V\left( \mathbf{r}  \right) \right) }{\partial x_{j}\partial x_{k}}\right\vert _{\mathbf{r}= \mathbf{0}}$ with $x_{j},x_{k} = x,y,z$  and $\hat{\mathbf{I}}^{2} = \hat{\mathbf{I}}_{x}^{2} + \hat{\mathbf{ I}}_{y}^{2}+\hat{\mathbf{I}}_{z}^{2}$. This Hamiltonian related at any set of principal axes system of coordinates satisfies the property of $V_{x_{j},x_{k}}=0$ for  $x_{j} \neq x_{k}$, and using the Laplace's equation $V_{x,x}+V_{y,y}+V_{z,z}=0$ then the Hamiltonian is rewritten as
\begin{equation}
\hat{\mathcal{H}}_{Q}=\frac{\hbar \omega _{Q}}{6 }\left( \left( 3 \hat{\mathbf{I}}_{z}^{2}-\hat{\mathbf{I}}^{2}\right) +\eta \left( \hat{  \mathbf{I}}_{x}^{2}-\hat{\mathbf{I}}_{y}^{2}\right) \right) \text{,}
\end{equation}%
where  $\ \omega _{Q}=\frac{eQV_{z,z}}{I\left( 2I-1\right) \hbar }$ defines  the quadrupolar angular frequency and  $\eta =\frac{V_{x,x}-V_{y,y}}{V_{z,z}}$ defines an asymmetry parameter.

The lyotropic liquid crystal used in this experimental implementation (see Fig. \ref{fig:ExperimentalDescription}(b)) follows a spatial arrangement with the molecules axis oriented along the strong static magnetic field. For this setup, the value of the asymmetry parameter is null, and the Hamiltonian of the quadrupolar contribution is
\begin{equation}
\hat{\mathcal{H}}_{Q}=\hbar \frac{\omega _{Q}}{6}\left( 3\hat{\mathbf{I}} _{z}^{2}-\hat{\mathbf{I}}^{2}\right)  \text{.}
\end{equation}
This Hamiltonian characterizes the quadrupolar energy contribution at the NMR Hamiltonian of Eq. (\ref{HamiltonianoRMN}) of the main text.


\begin{thebibliography}{99}

\bibitem{buzek1995Book} V. Bu\v{z}ek and P. L. Knight, \textit{Quantum interference, superposition states of light, and nonclassical effects}, Progress in Optics XXXIV (ed. E. Wolf) 1--158, Elsevier, Netherland (1995).

\bibitem{agarwal2013Book} G. S. Agarwal, \textit{Quantum Optics}, Cambridge, USA (2013).

\bibitem{monroe1996} C. Monroe, D. M.  Meekhof, B. E.  King and D. J. Wineland, 
{\href{https://science.sciencemag.org/content/272/5265/1131}{\textit{Science} \textbf{272}(5265), 1131--1136 (1996).}}

\bibitem{bouwmeester1997} Dik Bouwmeester, Jian-Wei  Pan, Klaus  Mattle, Manfred  Eibl, Harald  Weinfurter and Anton Zeilinger,  {\href{https://doi.org/10.1038/37539}{\textit{Nature} \textbf{390}(6660), 575--579 (1997).}}

\bibitem{leibfried2005} D. Leibfried, E.  Knill, S.  Seidelin, J.  Britton, R. B.  Blakestad, J.  Chiaverini, D. B.  Hume, W. M.  Itano, J. D.  Jost, C.  Langer, R.  Ozeri, R. Reichle and D. J. Wineland, {\href{http://dx.doi.org/10.1038/nature04251}{\textit{Nature} \textbf{438}(7068), 639--642 (2005).}}


\bibitem{deleglise2008} S. Deléglise, I. Dotsenko, C. Sayrin, J. Bernu, M. Brune, J.-M. Raimond and S. Haroche, {\href{https://doi.org/10.1038/nature07288}{\textit{Nature} \textbf{455}(7212), 510--514 (2008).}}

\bibitem{vlastakis2013} Brian Vlastakis, G. Kirchmair, Z. Leghtas, S. E. Nigg, L. Frunzio, S. M. Girvin, M. Mirrahimi, M. H. Devoret and   R. J. Schoelkopf, {\href{https://science.sciencemag.org/content/342/6158/607}{\textit{Science} \textbf{342}(6158), 607--610 (2013).}}

\bibitem{leek2013} Peter J. Leek, {\href{https://science.sciencemag.org/content/342/6158/568}{\textit{Science} \textbf{342}(6158), 568--569 (2013).}}

\bibitem{barreiro2011} J. T. Barreiro,  M. Müller, P. Schindler, D. Nigg, T. Monz, M. Chwalla, M. Hennrich, C. F. Roos,  P. Zoller,  and R. Blatt, {\href{https://doi.org/10.1038/nature09801}{\textit{Nature} \textbf{470}(7335), 486--491 (2011).}}

\bibitem{tiecke2014} T. G.  Tiecke, J. D. Thompson, N. P. de Leon, L. R. Liu, V. Vuletić and M. D. Lukin, {\href{https://doi.org/10.1038/nature13188}{\textit{Nature} \textbf{508}(7495), 241--244 (2014).}}

\bibitem{agarwal1997} G. S. Agarwal, R. R. Puri and R. P. Singh,, {\href{http://link.aps.org/doi/10.1103/PhysRevA.56.2249}{\textit{Phys. Rev. A} \textbf{56}(3), 2249--2254 (1997).}}

\bibitem{zheng2001} Shi-Biao Zheng, {\href{https://link.aps.org/doi/10.1103/PhysRevLett.87.230404}{\textit{Phys. Rev. Lett.} \textbf{87}(23), 230404 (2001).}}

\bibitem{klimov1998} A. B. Klimov and C. Saavedra {\href{http://www.sciencedirect.com/science/article/pii/S0375960198005295}{\textit{Physics Letters A} \textbf{247}(1), 14 - 20 (1998).}}

\bibitem{klimov2002JOB} A. B. Klimov and J. L. Romero and J. Delgado and L. L. S\'{a}nchez-Soto {\href{https://doi.org/10.1088%2F1464-4266%2F5%2F1%2F304}{\textit{Journal of Optics B: Quantum and Semiclassical Optics} \textbf{5}(1), 34--39 (2002).}}

\bibitem{james2000} Daniel F. V. James, {\href{https://onlinelibrary.wiley.com/doi/abs/10.1002/1521-3978%28200009%2948%3A9/11%3C823%3A%3AAID-PROP823%3E3.0.CO%3B2-M}{\textit{Fortschr. Phys.} \textbf{48}(9--11), 823 -- 837 (2000).}}

\bibitem{prado2011} F. O. Prado, F. S. Luiz, J. M. Villas-B\^oas, A. M. Alcalde, E. I. Duzzioni and L. Sanz,, {\href{https://link.aps.org/doi/10.1103/PhysRevA.84.053839}{\textit{Phys. Rev. A} \textbf{84}(5), 053839 (2011).}}

\bibitem{auccaise2015} R.  Auccaise, A. G. Araujo-Ferreira, R. S. Sarthour, I. S.  Oliveira, T. J. Bonagamba and I. Roditi, {\href{http://link.aps.org/doi/10.1103/PhysRevLett.114.043604}{\textit{Phys. Rev. Lett.} \textbf{114}(4), 043604 (2015).}}

\bibitem{gao2010} Wei-Bo  Gao, C.-Y.  Lu, X.-C.   Yao, P.   Xu, O.  Gühne, A.  Goebel, Y.-A.  Chen, C.-Z.  Peng, Z.-B.  Chen and J.-W. Pan, {\href{https://doi.org/10.1038/nphys1603}{\textit{Nature Physics} \textbf{6}(5), 331 - 335 (2010).}}

\bibitem{song2017PRL} C.  Song, K.  Xu, W.  Liu, C.-P.  Yang, S.-B.  Zheng, H.  Deng, Q. Xie, K.  Huang, Q.  Guo, L.  Zhang, P. Zhang,  Da Xu,  D. Zheng,  X. Zhu,  H. Wang,  Y.-A. Chen,  C.-Y. Lu,  S. Han and J.-W. Pan,, {\href{https://link.aps.org/doi/10.1103/PhysRevLett.119.180511}{\textit{Phys. Rev. Lett.} \textbf{119}(18), 180511 (2017).}}

\bibitem{cappellaro2005} P.  Cappellaro, J.  Emerson, N.  Boulant, C.  Ramanathan, S.  Lloyd and D. G. Cory, {\href{https://link.aps.org/doi/10.1103/PhysRevLett.94.020502}{\textit{Phys. Rev. Lett.} \textbf{94}(2), 020502 (2005).}}

\bibitem{slichter1992Book} Charles P. Slichter, \textit{Principles of magnetic resonance}, Springer International \textsc{T}hird enlarged edition (1992).

\bibitem{wasylishen2012Book} Roderick E. Wasylishen, Sharon E. Ashbrook, Stephen Wimperis, \textit{NMR of Quadrupolar Nuclei in Solid Materials}, John Wiley \& Sons Ltd. \textsc{F}irst edition (2012).

\bibitem{auccaise2009} R. Auccaise, J. Teles, T. J. Bonagamba, I. S. Oliveira, E. R. deAzevedo and R. S. Sarthour, {\href{http://link.aip.org/link/?JCP/130/144501/1}{\textit{The Journal of Chemical Physics} \textbf{130}(14), 144501 (2009).}}

\bibitem{auccaise2013} R. A. Estrada, E. R. de Azevedo, E. I. Duzzioni,  T. J. Bonagamba and M. H. Y. Moussa, {\href{http://dx.doi.org/10.1140/epjd/e2013-30689-1}{\textit{Eur. Phys. J. D} \textbf{67}(6), 127 (2013).}}

\bibitem{araujo-ferreira2013} A. G. Araujo-Ferreira, R.  Auccaise, R. S.  Sarthour,  I. S. Oliveira, T. J.  Bonagamba  and  I. Roditi, {\href{http://link.aps.org/doi/10.1103/PhysRevA.87.053605}{\textit{Phys. Rev. A} \textbf{87}(5), 053605 (2013).}}

\bibitem{nie2015} X. Nie, J. Li, J. Cui, Z. Luo, J. Huang, H. Chen, C. Lee, X. Peng and J.-F. Du, {\href{http://stacks.iop.org/1367-2630/17/i=5/a=053028}{\textit{New Journal of Physics} \textbf{17}(5), 053028 (2015).}}

\bibitem{teles2018} J. Teles, R. Auccaise, C. Rivera-Ascona, A. G. Araujo-Ferreira, J. P. Andreeta and T. J. Bonagamba,, {\href{https://doi.org/10.1007/s11128-018-1947-1}{\textit{Quantum Information Processing} \textbf{17}(7), 177 (2018).}}

\bibitem{oliveira2007Book} I. S. Oliveira, T. J.  Bonagamba,  R. S. Sarthour, J. C. C.  Freitas  and  E. R. deAzevedo, \textsc{N}\textsc{M}\textsc{R}  \textit{Quantum Information Processing}, Elsevier - Amsterdan (2007).

\bibitem{teles2007} J. Teles, E. R. deAzevedo, R. Auccaise, R. S. Sarthour, I. S. Oliveira and T. J. Bonagamba, {\href{http://link.aip.org/link/?JCP/126/154506/1}{\textit{The Journal of Chemical Physics} \textbf{126}(15), 154506 (2007).}}

\bibitem{perelomov1985Book} A. Perelomov, \textit{Generalized Coherent States and Their Applications}, Text and Monographs in Physics - Springer-Verlag (1985).

\bibitem{fortunato2002}  E. M. Fortunato,  M. A. Pravia, N.  Boulant, G.  Teklemariam,   T. F.  Havel, and  D. G. Cory,  {\href{https://doi.org/10.1063/1.1465412}{\textit{The Journal of Chemical Physics} \textbf{116}(17), 7599-7606 (2002).}}

\bibitem{agarwal1981} G. S. Agarwal,  {\href{http://link.aps.org/doi/10.1103/PhysRevA.24.2889}{\textit{Phys. Rev. A} \textbf{24}(6), 2889--2896 (1981).}}

\bibitem{benedict1999}  M. G. Benedict and  A. Czirj\'ak, {\href{http://link.aps.org/doi/10.1103/PhysRevA.60.4034}{\textit{Phys. Rev. A} \textbf{60}(5), 4034--4044 (1999).}}

\bibitem{sanchez-soto2013} L L S\'{a}nchez-Soto, A. B. Klimov, P. de la Hoz and G Leuchs, {\href{http://stacks.iop.org/0953-4075/46/i=10/a=104011}{\textit{Journal of Physics B: Atomic, Molecular and Optical Physics} \textbf{46}(10), 104011 (2013).}}

\bibitem{ourjoumtsev2006} A.  Ourjoumtsev, R.  Tualle-Brouri, J.  Laurat and P. Grangier,  {\href{https://science.sciencemag.org/content/312/5770/83}{\textit{Science} \textbf{312}(5770), 83--86 (2006).}}

\bibitem{ourjoumtsev2007} A. Ourjoumtsev, H.  Jeong, R.  Tualle-Brouri and P. Grangier,  {\href{http://dx.doi.org/10.1038/nature06054}{\textit{Nature} \textbf{448}(7155), 784--786 (2007).}}

\bibitem{leibfried1996} D.  Leibfried, D. M.  Meekhof, B. E.  King, C.  Monroe, W. M.  Itano and D. J. Wineland,  {\href{https://link.aps.org/doi/10.1103/PhysRevLett.77.4281}{\textit{Phys. Rev. Lett.} \textbf{77}(21), 4281--4285 (1996).}}

\bibitem{teles2015} J. Teles, C. Rivera-Ascona, R. S. Polli, R. Oliveira-Silva, E. L. G. Vidoto, J. P. Andreeta and T. J. Bonagamba,  {\href{http://dx.doi.org/10.1007/s11128-015-0967-3}{\textit{Quantum Information Processing} \textbf{14}(6), 1889--1906 (2015).}}

\bibitem{garon2015} A.  Garon, R.  Zeier and S. J. Glaser,  {\href{https://link.aps.org/doi/10.1103/PhysRevA.91.042122}{\textit{Phys. Rev. A} \textbf{91}(4), 042122 (2015).}}

\bibitem{koczor2019JPA} B. Koczor, R. Zeier and S. J. Glaser,  {\href{https://doi.org/10.1088%2F1751-8121%2Faaf302}{\textit{Journal of Physics A: Mathematical and Theoretical} \textbf{52}(5), 055302 (2019).}}

\bibitem{koczor2019AP} B. Koczor, R. Zeier and S. J. Glaser,  {\href{http://www.sciencedirect.com/science/article/pii/S0003491618303117}{\textit{Annals of Physics} \textbf{408}, 1 - 50 (2019).}}

\bibitem{neergaard-nielsen2006} J. S.  Neergaard-Nielsen, B. M.  Nielsen, C. Hettich, K.  M\o{}lmer and E. S. Polzik,  {\href{https://link.aps.org/doi/10.1103/PhysRevLett.97.083604}{\textit{Phys. Rev. Lett.} \textbf{97}(8), 083604 (2006).}}

\bibitem{lucke2014} B.  L\"ucke, J. Peise, G.  Vitagliano, J. Arlt, L.  Santos, G.  T\'oth and C. Klempt,  {\href{http://dx.doi.org/10.1038/nature06054}{\textit{Phys. Rev. Lett.} \textbf{112}(15), 155304 (2014).}}

\bibitem{kampermann2005} H. Kampermann  and W. S. Veeman,  {\href{https://doi.org/10.1063/1.1904595}{\textit{The Journal of Chemical Physics} \textbf{122}(21), 214108 (2005).}}

\bibitem{yusa2005} G.  Yusa, K.  Muraki, K.   Takashina, K.  Hashimoto and Y. Hirayama,,  {\href{http://dx.doi.org/10.1038/nature03456}{\textit{Nature} \textbf{434}(7036), 1001--1005 (2005).}}

\bibitem{miranowicz2015} A.  Miranowicz, \ifmmode \mbox{\c{S}}\else \c{S}. K.  \"Ozdemir, J.  Bajer, G.  Yusa, N.  Imoto, Y. Hirayama and F. Nori,,  {\href{https://link.aps.org/doi/10.1103/PhysRevB.92.075312}{\textit{Phys. Rev. B} \textbf{92}(7), 075312 (2015).}}

\bibitem{hendrickx2021} N. W.  Hendrickx, W. I. L.   Lawrie, M.  Russ, F.  van Riggelen, S. L.  de Snoo, R. N.  Schouten, A.   Sammak, G.  Scappucci and M. Veldhorst,  {\href{https://doi.org/10.1038/s41586-021-03332-6}{\textit{Nature} \textbf{591}(7851), 580-585 (2021).}}

\bibitem{glenn2018} D. R.  Glenn, D. B. Bucher, J.  Lee, M. D.   Lukin, H. Park and R. L. Walsworth,  {\href{https://doi.org/10.1038/nature25781}{\textit{Nature} \textbf{555}(7696), 351-354 (2018).}}

\bibitem{dutt2007} M. V. G.  Dutt, L.  Childress, L.  Jiang, E.  Togan, J.  Maze, F.  Jelezko, A. S.  Zibrov, P. R.  Hemmer and M. D. Lukin,  {\href{http://www.sciencemag.org/content/316/5829/1312.abstract}{\textit{Science} \textbf{316}(5829), 1312-1316 (2007).}}

\bibitem{kong2016} F.  Kong, C.  Ju, Y.  Liu, C.  Lei, M.  Wang, X.  Kong, P.-F.  Wang, P.  Huang, Z.  Li, F. Shi, L.  Jiang and J.-F. Du,  {\href{http://link.aps.org/doi/10.1103/PhysRevLett.117.060503}{\textit{Phys. Rev. Lett.} \textbf{117}(6), 060503 (2016).}}

\bibitem{rose2018} B. C.  Rose, D.  Huang, Z.-H.  Zhang, P.  Stevenson, A. M.  Tyryshkin, S.  Sangtawesin, S.  Srinivasan, L. Loudin, M. L.  Markham, A. M.  Edmonds, D. J.  Twitchen, S. A. Lyon and N. P. de Leon,,  {\href{http://science.sciencemag.org/content/361/6397/60}{\textit{Science} \textbf{361}(6397), 60--63 (2018).}}

\bibitem{aslam2017} N.  Aslam, M.  Pfender, P.  Neumann, R.  Reuter, A.  Zappe, F.  F. de Oliveira, A.  Denisenko, H.  Sumiya, S.  Onoda, J.  Isoya and J. Wrachtrup,  {\href{https://www.science.org/doi/abs/10.1126/science.aam8697}{\textit{Science} \textbf{357}(6346), 67-71 (2017).}}

\bibitem{evans2018} R. E.  Evans, M. K. Bhaskar, D. D.  Sukachev, C. T.  Nguyen, A.  Sipahigil, M. J.  Burek, B.  Machielse, G. H.  Zhang, A. S.  Zibrov, E.  Bielejec, H.  Park, M.  Lon{\v c}ar  and M. D. Lukin,  {\href{http://science.sciencemag.org/content/362/6415/662}{\textit{Science} \textbf{362}(6415), 662--665 (2018).}}

\bibitem{lesik2019} M. Lesik, T. Plisson, L. Toraille, J. Renaud, F. Occelli, M. Schmidt, O. Salord, A. Delobbe, Th. Debuisschert, L. Rondin, P. Loubeyre and J.-F. Roch,  {\href{https://science.sciencemag.org/content/366/6471/1359}{\textit{Science} \textbf{366}(6471), 1359--1362 (2019).}}

\bibitem{das2003} R.   Das and A. Kumar, {\href{https://link.aps.org/doi/10.1103/PhysRevA.68.032304}{\textit{Phys. Rev. A} \textbf{68}(3), 032304 (2003).}}

\bibitem{hacker2019} B. Hacker, S. Welte, S. Daiss, A. Shaukat, S. Ritter, L. Li and  G. Rempe,  {\href{https://doi.org/10.1038/s41566-018-0339-5}{\textit{Nature Photonics} \textbf{13}(2), 110 -- 115 (2019).}}

\bibitem{omran2019} A.  Omran, H.  Levine, A.  Keesling, G.  Semeghini, T. T.  Wang, S.  Ebadi, H.  Bernien, A. S.  Zibrov, H.  Pichler, S.  Choi, J.  Cui, M.  Rossignolo, P.  Rembold, S.  Montangero, T.  Calarco, M.  Endres, M.  Greiner, V.  Vuleti{\'c} and M. D. Lukin,  {\href{https://science.sciencemag.org/content/365/6453/570}{\textit{Science} \textbf{365}(6453), 570--574 (2019).}}

\bibitem{auccaise2008} R. Auccaise, J. Teles, R.S. Sarthour, T.J. Bonagamba, I.S. Oliveira and E.R. deAzevedo,  {\href{http://www.sciencedirect.com/science/article/B6WJX-4RPTJ8M-1/2/a2ff51926c5e3feb296e0b09c8f3bbfe}{\textit{Journal of Magnetic Resonance} \textbf{192}(1), 17 - 26 (2008).}}

\bibitem{quist1992} P.-O.   Quist, B.   Halle and I.  Fur\'o,  {\href{https://doi.org/10.1063/1.461892}{\textit{The Journal of Chemical Physics} \textbf{96}(5), 3875--3891 (1992).}}

\bibitem{villamizar2018} D. V.  Villamizar, E. I.  Duzzioni, A. C. S. Leal  and R. Auccaise,  {\href{https://link.aps.org/doi/10.1103/PhysRevA.97.052125}{\textit{Phys. Rev. A} \textbf{97}(5), 052125 (2018).}}

\bibitem{varshalovich1988Book} D. A. Varshalovich, A. N. Moskalev and V. K. Khersonskii, \textit{Quantum Theory of Angular Momentum}, World Scientific Publishing Co. Pte. Ltd - A Editora (1988).

\bibitem{jin2007PRA} G.-R.  Jin  and  S. W. Kim,  {\href{https://link.aps.org/doi/10.1103/PhysRevA.76.043621}{\textit{Phys. Rev. A} \textbf{76}(4), 043621 (2007).}}

\bibitem{gati2007} R. Gati and M. K. Oberthaler,
{\href{http://iopscience.iop.org/0953-4075/40/10/R01}{\textit{Journal of Physics B} \textbf{40}, R61--R89 (2007).}}

\bibitem{cohen1977Book} Cohen-Tannoudji, C. and Bernard Diu and Franck Laloë,
{{\textit{Quantum Mechanics} New York : Wiley (1977).}}






\end{thebibliography}
\end{document}